%% file: main.tex
\documentclass[sigconf]{acmart}

\usepackage{subcaption}
\usepackage{soul}
\usepackage{xspace}
\usepackage{multirow}
\usepackage{xurl}
\usepackage{url}
\usepackage{tcolorbox}
\usepackage{enumitem}
\usepackage{hyperref}
\usepackage[frozencache,cachedir=.]{minted}

\copyrightyear{2024}
\acmYear{2024}
\setcopyright{rightsretained}
\acmConference[ICSE '24]{2024 IEEE/ACM 46th International Conference on Software Engineering}{April 14--20, 2024}{Lisbon, Portugal}
\acmBooktitle{2024 IEEE/ACM 46th International Conference on Software Engineering (ICSE '24), April 14--20, 2024, Lisbon, Portugal}\acmDOI{10.1145/3597503.3623342}
\acmISBN{979-8-4007-0217-4/24/04}

\newcommand{\tool}{\textit{LLMAO}\xspace}
\newcommand{\bpy}{\textit{BugsInPy}\xspace}
\newcommand{\devign}{\textit{Devign}\xspace}
\newcommand{\defj}{\textit{Defects4J}\xspace}
\newcommand{\codegen}{\textit{CodeGen}\xspace}
\definecolor{dawnblue}{rgb}{0.84, 0.92, 1.0}

\definecolor{LightGray}{gray}{0.9}

\begin{document}
\title{Large Language Models for Test-Free Fault Localization}
\author{Aidan Z.H. Yang}
\email{aidan@cmu.edu}
\affiliation{
\institution{Carnegie Mellon University}
\city{Pittsburgh}
\country{United States}
}

\author{Claire Le Goues}
\email{clegoues@cs.cmu.edu}
\affiliation{
\institution{Carnegie Mellon University}
\city{Pittsburgh}
\country{United States}
}

\author{Ruben Martins}
\email{rubenm@cs.cmu.edu}
\affiliation{
\institution{Carnegie Mellon University}
\city{Pittsburgh}
\country{United States}
}

\author{Vincent J. Hellendoorn}
\email{vhellendoorn@cmu.edu}
\affiliation{
\institution{Carnegie Mellon University}
\city{Pittsburgh}
\country{United States}
}

\begin{abstract}
Fault Localization (FL) aims to automatically localize buggy lines of code, a key first step in many manual and automatic debugging tasks. Previous FL techniques assume the provision of input tests, and often require extensive program analysis, program instrumentation, or data preprocessing.  Prior work on deep learning for APR struggles to learn from small datasets and produces limited results on real-world programs. Inspired by the ability of large language models (LLMs) of code to adapt to new tasks based on very few examples, we investigate the applicability of LLMs to line level fault localization. Specifically, we propose to overcome the left-to-right nature of LLMs by fine-tuning a small set of bidirectional \emph{adapter} layers on top of the representations learned by LLMs to produce \tool, the first language model based fault localization approach that locates buggy lines of code without any test coverage information.
We fine-tune LLMs with 350 million, 6 billion, and 16 billion parameters on small, manually curated corpora of buggy programs such as the \defj corpus. We observe that our technique achieves substantially more confidence in fault localization when built on the larger models, with bug localization performance scaling consistently with the LLM size.
Our empirical evaluation shows that \tool improves the Top-1 results over the state-of-the-art machine learning fault localization (MLFL) baselines by 2.3\%-54.4\%, and Top-5 results by 14.4\%-35.6\%. \tool is also the first FL technique trained using a language model architecture that can detect security vulnerabilities down to the code line level.
\end{abstract}

\ccsdesc[500]{Software and its engineering~Software functional properties}
\ccsdesc[500]{Computing methodologies~Neural networks}

\maketitle
\input{sections/introduction}

\input{sections/motivation}
\input{sections/approach}
\input{sections/evaluation}
\input{sections/related_work}
\input{sections/threats}
\input{sections/conclusion}

\section*{Acknowledgements}
This work was partially supported by the US National Science Foundation (NSF) awards CCF-1750116 and CCF-1762363, and by 
ANI 045917 award funded by FEDER and Portuguese Foundation
for Science and Technology (FCT).

\bibliographystyle{ieeetr}
\bibliography{reference}

\end{document}

%% file: sections/introduction.tex
\section{Introduction}
\label{sec:intro}

Fault localization (FL)~\cite{abreu2007accuracy,li2019deepfl,lou2021boosting,li2022fault} approaches aim to automatically identify which program entities (like a line, statement, module, or file) are implicated in a particular bug. The goal is to assist programmers in fixing defects by pinpointing the places in the code base that should be modified to fix them. 

Broadly speaking, existing FL techniques combine or leverage
static and dynamic program analysis information to compute a score corresponding to a program entity's probability of contributing to a particular bug. 
\textit{Spectrum based fault localization (SBFL)} approaches, such as Tarantula \cite{jones2005empirical} or Ochiai \cite{abreu2006evaluation}, apply statistical analysis on the coverage data of failed/passed tests to compute the suspiciousness of code elements. SBFL relies exclusively on test coverage and is thus less applicable for data-driven defects; it is also  sensitive to properties of the underlying test suite like coverage and numbers of passing and failing tests~\cite{abreu2007accuracy}.  
\textit{Mutation based fault localization (MBFL)} (like FIFL~\cite{zhang2013injecting} or Metallaxis~\cite{moon2014ask}) also analyzes test case behavior to localize faults, but uses mutation analysis to assess the concrete impact of particular code lines on test outcomes. While effective, MBFL approaches are computationally intensive and their performance is highly variable~\cite{chekamassessing}. 

Recent advances in \textit{Machine learning based fault localization (MLFL)}, like DeepFL~\cite{li2019deepfl}, DEEPRL4FL~\cite{li2021fault}, and GRACE~\cite{lou2021boosting} 

use machine learning to relate code, test, or execution features to the likelihood of faultiness for a program entity.  MLFL techniques learn to detect faulty lines of code from information including suspiciousness scores from existing SBFL and MBFL techniques (e.g., TRANSFERFL~\cite{meng2022improving}),  
fault-proneness features like code complexity metrics (e.g., DeepFL~\cite{li2019deepfl}), or
the test coverage matrix (DEEPRL4FL~\cite{li2021fault}), among others.
These approaches speak to the potential that increasingly powerful machine learning models have for supporting debugging tasks. 

Indeed, Deep learning (DL) has shown promise for many code related tasks, such as program synthesis ~\cite{vaithilingam2022expectation, nijkamp2022codegen}. The most effective DL models for both natural language and code related tasks are large language models (LLMs), such as Codex~\cite{chen2021evaluating} and GPT-4~\cite{openai2023gpt4}.
This class of models trains many billions of parameters with even more tokens of training data, which tends to yield highly flexible and powerful text generators.
LLMs' utility for code generation and the fact that they are trained on an abundance of publicly-available code~\cite{chen2021evaluating} both suggest that existing large-scale LLMs capture program source code in ways that can be leveraged for specialized development tasks.
A key property of LLMs is that their performance improves consistently with the \emph{scale} of their computational budget \cite{kaplan2020scaling}, which is itself a function of the model and training data size.

For instance, LLM performance on program synthesis benchmarks increases linearly with the magnitude (log scale) of the number of parameters in the model~\cite{xu2022systematic}. This suggests that there is substantial performance to be unlocked for software engineering tasks by leveraging the largest publicly available language models. However, most existing work in this space to date either trains small models from scratch~\cite{li2019deepfl, li2021fault, li2022dear}, or fine-tunes modest-sized models~\cite{meng2022improving}, missing out on the scale of state-of-the-art LLMs. 

This is in part because LLMs are not immediately suited off-the-shelf for coding tasks that do not involve code generation, like fault localization. State-of-the-art LLMs for code~\cite{chen2021evaluating, black2021gpt, tunstall2022natural, nijkamp2022codegen} are trained to generate code in a left-to-right manner, with each token predicted from its preceding context. 
We posit that models trained in this way are less suitable for token-level discriminative tasks, like line-level fault localization, because the representation for any given token is only conditioned on the context to the left. 

In this paper, we present a promising alternative: we train lightweight bidirectional \emph{adapters}, themselves small models of the same architecture as the base LLM, on top of left-to-right language models. These adapters add relatively few parameters and can be trained effectively on small datasets of real bugs, such as \defj~\cite{just2014defects4j}, without updating the underlying LLM. We demonstrate that the representations learned by pretrained left-to-right language models already contain a wealth of knowledge about the suspiciousness of lines of code, which increases strongly with the size of the LLM.  We can leverage this power through our adapters to find bugs while requiring just a few hundred training samples for pretraining. 
Our approach yields better fault localization performance than prior work while requiring significantly less data preprocessing overhead. Importantly, our approach does not use test cases at all, and therefore does not depend on test code quality for its performance. 
Our approach does not need to run or analyze the test cases or examine the program behavior on test cases to perform localization.
Because the approach is lightweight, it can effectively fine-tune existing LLMs for particular languages (we show applicability to C, Java, and Python), or particular defect classes (we show applicability to functional defects and security vulnerabilities), with a relatively small amount of training data.

In summary, we make the following contributions.
\begin{itemize}

\item{\textbf{\tool.} We propose a technique that uses different configurations of language models to predict faulty lines across three languages and two different application domains.}
\item{\textbf{Novel large language model based learning for FL}. We showed that with fine-tuning on top of off-the-shelf large language models, we can achieve a higher fault detection rate than previous MLFL techniques without the use of test cases.}
\item{\textbf{DL based security vulnerability detection}. \tool is the first MLFL technique that can detect code line level vulnerabilities in the security domain.}
\item{\textbf{Empirical evaluation}. We evaluated \tool against recent state-of-the-art FL models to show its effectiveness in fault localization.}
\item{\textbf{Artifact availability}}. Our data, tool, and model checkpoints are available.\footnote{\url{https://github.com/squaresLab/LLMAO}}
\end{itemize}

%% file: sections/motivation.tex
\section{Motivation}
\label{sec:motivation}

In this section, we discuss in detail two real-world bugs that test-based FL techniques struggle to clearly localize.  We use these examples to motivate why we propose a novel language model based fault localization technique that shifts the dependence on tests to an LLM's latent understanding of source code.

\subsection{General Logic Defects}
\label{sec:motivation-general}
Consider Figure~\ref{code:lang47-code}, which shows snippets of code from two methods in the Apache Commons Lang project. 
Lang-47 (i.e., bug \#47 of the Lang project) from the \defj (V1.2.0)~\cite{just2014defects4j} dataset highlights a null pointer exception that can be triggered in both of these methods, for the same reason.  The issue was addressed by adding the null pointer check and initialization shown starting on line 7 in \texttt{appendFixedWidthPadRight}; the identical code and block added in \texttt{appendFixedWidthPadLeft} is elided for brevity. 

\begin{figure}
\centering
\begin{subfigure}[t]{0.48\textwidth}

\begin{minted}[
frame=lines,
framesep=2mm,
baselinestretch=1.2,
fontsize=\footnotesize,
linenos,
highlightlines={7},
highlightcolor=pink,
]{Java}
public StrBuilder appendFixedWidthPadRight(Object, int, char) {
  ...
  if (width > 0) { 
    ensureCapacity(size + width); // SBFL=0.35
    String str = (obj == null ? getNullText() 
        : obj.toString()); //SBFL=0.35
    int strLen = str.length(); //SBFL=0.35
    ...
public StrBuilder appendFixedWidthPadLeft(Object, int, char) {
 // relevant code identical to the above
    ...
public String getNullText(){ 
  return nullText;  // SBFL=0.71
}
\end{minted}
\caption{\small Code snippet implicated in Apache Commons Lang bug \#47 from \defj. Both methods throw a \texttt{NullPointerException} when \texttt{getNullText()} also returns \texttt{null} (line 7).  The developer addressed this by adding null checks after the assignment to \texttt{str} (not shown).
Select lines are annotated with Ochiai~\cite{abreu2006evaluation} suspiciousness score.}
\label{code:lang47-code}
\end{subfigure}

\begin{subfigure}[t]{0.48\textwidth}
\begin{minted}[
frame=lines,
framesep=2mm,
baselinestretch=1.2,
fontsize=\footnotesize,
linenos,
highlightlines={},
highlightcolor=pink
]{Java}
public void testLang412Right() { 
    StrBuilder sb = new StrBuilder();
    sb.appendFixedWidthPadRight(null, 10, '*');
    assertEquals( "Failed to invoke appendFixedWidthPadRight", 
    "**********", sb.toString());
} //Test fails due to NullPointerException in appendFixedWidthPadRight()
public void testLang412Left() { 
    StrBuilder sb = new StrBuilder();
    sb.appendFixedWidthPadLeft(null, 10, '*');
    assertEquals( "Failed to invoke appendFixedWidthPadLeft", 
    "**********", sb.toString());
} //Test fails due to NullPointerException in appendFixedWidthPadLeft()
\end{minted}
\caption{\small Lang's bug \#47 and corresponding failed tests}
\label{code:lang47-tests}
\end{subfigure}
\caption{\small Apache Commons Lang Bug \#47, from \defj}
\label{code:lang47}
\end{figure}

Given tests, we can use the Ochiai SBFL formula~\cite{abreu2006evaluation} to calculate code line suspiciousness scores to help pinpoint this bug. 
SBFL techniques in general compute suspiciousness by applying a formula to each entity (line, in this case) in the codebase based on test coverage information for passing and failing tests. 
Specifically, Ochiai counts for each code line $(\ell)$ the number of failed tests covering $\ell$ $(\ell_f)$ or not covering $\ell$ $(n_f)$, and the number of passed tests covering $\ell$ $(\ell_p)$ or not covering $\ell$ $(n_p)$.
The suspiciousness score of a code line $(\ell)$ is then 
$Ochiai(\ell) = \ell_f(\ell_f + \ell_p)^{(-\frac{1}{2})}(\ell_f + n_f)^{(-\frac{1}{2})}$.

Several tests in the Apache Commons Lang test suite execute this code.  
The two that throw null pointer exceptions, demonstrating the bug, are shown in Figure~\ref{code:lang47-tests}.  Five others (not shown) execute these two methods as well, but are passing. 

Figure~\ref{code:lang47-code} shows Ochiai scores computed using these tests. 
The scores demonstrate a common limitation of SBFL, which is that it cannot disambiguate between lines in a single straight-line block, as shown in \texttt{appendFixedWidthPadRight}.  \texttt{testLang412Right()} executes lines 1--7, corresponding to the \texttt{then} block of the check on line 3. 
This computation is also misled by the small number of triggering tests: both failing tests cover
\texttt{getNullText}, while only two of the five passing tests do.
Line 13 in Figure~\ref{code:lang47-code} has a much higher score than the code in the two methods that call it.\footnote{Note that the test suite includes another test, \texttt{testGetSetNullText}, which explicitly checks that \texttt{getNullText} returns \texttt{null} (not the empty string).}

MLFL techniques like DeepFL~\cite{li2019deepfl} use other features on top of SBFL suspiciousness scores for training data, like textual similarity information,
to guide their model to detect faulty methods. 
However, DeepFL only has confidence in method-level fault localization, with limited results at the statement level.

Our technique can detect line 7 from Figure \ref{code:lang47-code} as highly suspicious. It assigned a score of 0.33 on line 7, and ranks it the fourth most suspicious line in the code file. Our technique also assigned a score of 0.27 on line 1, and ranks it the seventh most suspicious line in the code file. In contrast, our technique only assigned a score of 0.09 for line 13, which is not within the top 20.
Language models are good at detecting these types of defects because they recognize unlikely inputs \cite{ray2016naturalness}. Consider just the text of the code, line 5 appears to assign a \texttt{null}-like value (the result of \texttt{getNullText()}) to \texttt{src} under some conditions. Line 7 then invokes a method on \texttt{src}. Even without knowing the implementation of \texttt{getNullText()} in depth (for which traditional program analyzers would be more suitable than language models), this pattern is suspicious to a human reader and a large language model alike.

\begin{figure}[t]
\centering
\begin{minted}[
frame=lines,
framesep=2mm,
baselinestretch=1.2,
fontsize=\footnotesize,
linenos,
highlightlines={10,12, 23-25},
highlightcolor=pink
]{C}
DISAS_INSN(divw)
{
    TCGv reg;
    TCGv tmp;
    TCGv src;
    int sign;
    sign = (insn & 0x100) != 0;
    reg = DREG(insn, 9);
    if (sign) {
        tcg_gen_ext16s_i32(QREG_DIV1, reg);
    } else {
        tcg_gen_ext16u_i32(QREG_DIV1, reg);
    }
    SRC_EA(env, src, OS_WORD, sign, NULL);
    tcg_gen_mov_i32(QREG_DIV2, src);
    if (sign) {
        gen_helper_divs(cpu_env, tcg_const_i32(1));
    } else {
        gen_helper_divu(cpu_env, tcg_const_i32(1));
    }
    tmp = tcg_temp_new();
    src = tcg_temp_new();
    tcg_gen_ext16u_i32(tmp, QREG_DIV1);
    tcg_gen_shli_i32(src, QREG_DIV2, 16);
    tcg_gen_or_i32(reg, tmp, src);
    set_cc_op(s, CC_OP_FLAGS);
}
\end{minted}
\caption{\small Qemu's CWE-362 (Race condition vulnerability)}
\label{code:qemu}
\end{figure}

\subsection{Vulnerability Detection}

Logical errors are not the only type of code mistakes that can impact software quality.  
Software security vulnerabilities are often the target of various forms of cyber-attacks.

The \devign dataset~\cite{zhou2019devign} labels vulnerable functions from four open-source C-language repositories (requiring 600 man hours of manual labeling). 
Figure \ref{code:qemu} shows a bug from the Qemu open-source project,\footnote{https://qemu.org/}  one of the four studied repositories in \devign. The bug lines (highlighted) correspond to CWE-362, within the top 25 most dangerous Common Weakness Enumeration (CWE) list.\footnote{https://cwe.mitre.org//top25/archive/2022/2022\_cwe\_top25.html}
CWE describes CWE-362 as concurrent execution using shared resource with improper synchronization (i.e., race condition).
Although Qemu's repository\footnote{https://github.com/qemu/qemu} includes test cases for crashes and input behaviors, none of the test cases covers concurrency bugs that only occur during run time. 
Indeed, concurrency bugs like race conditions are ill-suited for discovery via traditional testing.

As a result, test based fault localization and debugging methods are clearly inapplicable to this kind of defect.  
This has of course motivated
significant work in profiling and analysis to discover and address them~\cite{zhou2019devign, chakraborty2021deep}. Chakraborty et al. \cite{chakraborty2021deep} found that existing modeling techniques do not completely capture code semantics in vulnerability detection. Existing deep learning based vulnerability detection tools only go as far as predicting any vulnerability in a code snippet or program file, rather than individual statements.
Traditional approaches such as static analysis can be used to detect race conditions~\cite{DBLP:conf/sosp/EnglerA03,DBLP:conf/sigsoft/VoungJL07}. However, these approaches are either precise but not scalable or can scale for large programs but incur a high false positive rate, limiting their usage in practice. 

Fortunately, a dataset like \devign encompasses significant manual effort in labeling existing security vulnerabilities in existing code, as has been done for lines 10, 12, and 23--25 in Figure~\ref{code:qemu}.  

We show in this paper that an FL-specific model pretrained on a large-scale LLM can also detect security vulnerabilities without test cases. 

Our technique detects lines 3, 4, 10, 12, and 23 as faulty in Figure \ref{code:qemu}, in which lines 10, 12, and 23-25 are actual vulnerability lines. Our technique successfully localizes three of the five lines that are faulty. Surprisingly, lines 3 and 4 are variable declarations (i.e., variables \texttt{reg} and \texttt{tmp}) for the actual faulty lines 10--12, and 23--25.

%% file: sections/approach.tex
\begin{figure*}
\centering
\includegraphics[width=\textwidth]{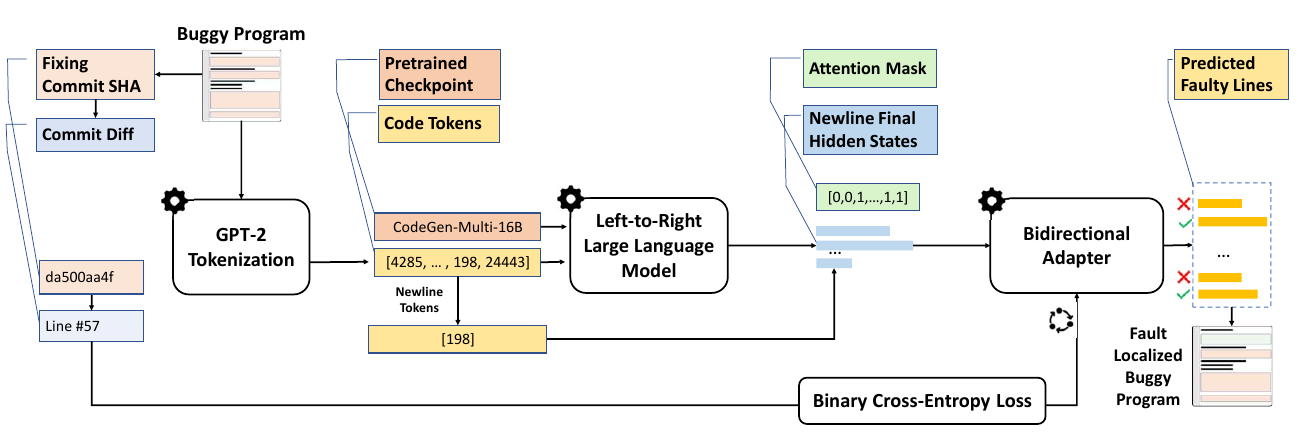}
\caption{\small \tool’s architecture, which takes as input a buggy program and produces a list of suspiciousness scores for each code line}
\label{fig:overview}
\end{figure*}

\begin{figure}
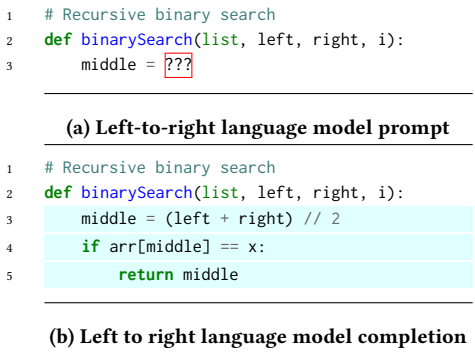

\centering
\begin{subfigure}[t]{0.32\textwidth}
\begin{minted}[
frame=lines,
framesep=2mm,
baselinestretch=1.2,
fontsize=\footnotesize,
linenos
]{Python}
# Recursive binary search
def binarySearch(list, left, right, i):
    middle = ???
\end{minted}
\caption{\small Left-to-right language model prompt}
\label{code:prompt}
 \end{subfigure}
\begin{subfigure}[t]{0.32\textwidth}
\begin{minted}[
frame=lines,
framesep=2mm,
baselinestretch=1.2,
fontsize=\footnotesize,
linenos,
highlightlines={3-5}
]{Python}
# Recursive binary search
def binarySearch(list, left, right, i):
    middle = (left + right) // 2
    if arr[middle] == x:
        return middle
\end{minted}
\caption{\small Left to right language model completion}
\label{code:completion}
 \end{subfigure}
 \caption{\small Left-to-right language model code generation}
 \label{code:leftright}
\end{figure}

\section{Approach}
\label{sec:approach}
In this section, we discuss the key ideas behind our language model enabled fault localization technique.
Figure \ref{fig:overview} shows an overview of \tool's training setup. The input to \tool is a buggy program; its output is a list of \emph{suspiciousness} scores corresponding to each code line's probability of being buggy -- values close to 1 indicate that lines are likely defective. As shown in Figure \ref{fig:overview}, we first tokenize the input and then provide it to a pretrained left-to-right LLM. From this LLM, we obtain one (high-dimensional) \emph{vector representation} per line, which we provide to a small bidirectional model that predicts bugginess probabilities for each line. We only train the final stage of this model; the LLM remains frozen and can be easily replaced with other powerful open-source models. Figure \ref{fig:attention} shows a more detailed description of our language modeling procedure, which we describe in detail in Section \ref{sec:adapter}.
In the following sections, we describe each component of \tool.

\subsection{Left-to-right Language Models}
\begin{figure}[t]
\centering
\includegraphics[width=0.49\textwidth]{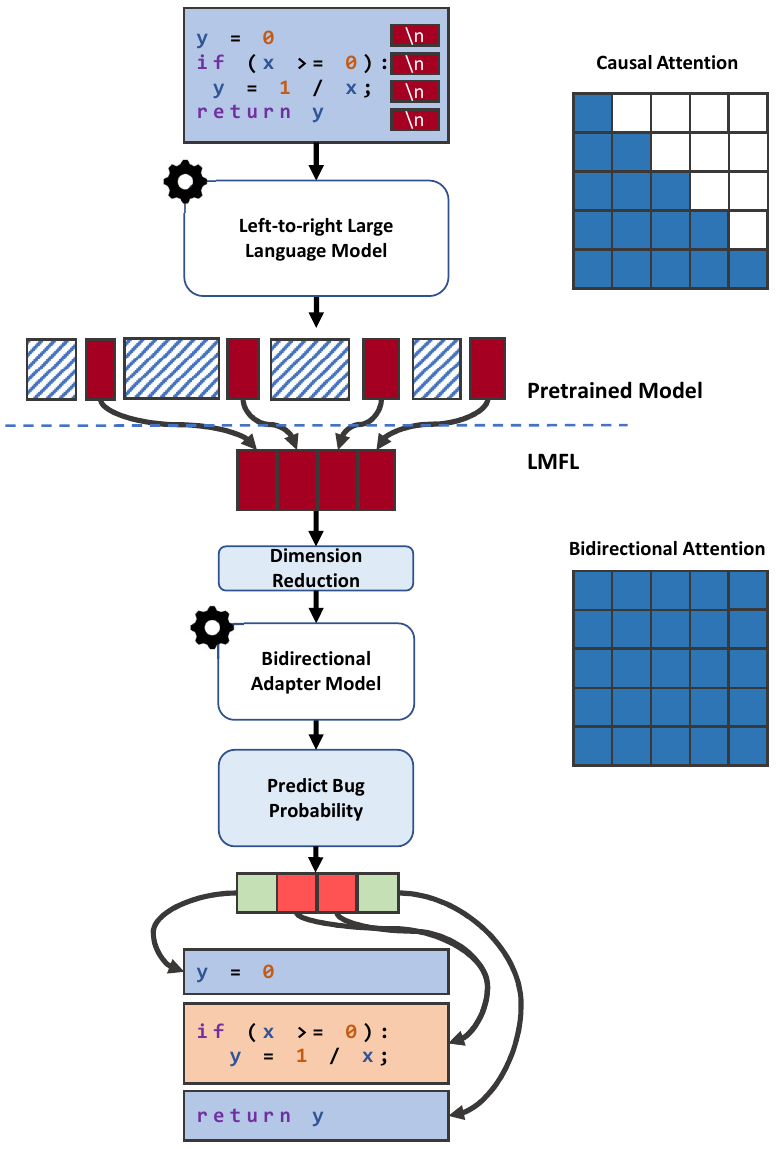}
\caption{\small Attention masking procedure of \tool}
\label{fig:attention}
\end{figure}

Neural Language Models typically produce text in a left-to-right manner, producing each word given its prefix context. This both enables efficient training, as any document can be turned into a collection of as many training samples as there are tokens, and enables them to generate new text once trained.
Virtually all modern language models are attention-based models that use the Transformer architecture \cite{vaswani2017attention}. In these models, each token exchanges information with all other tokens via a learned attention procedure. To efficiently train left-to-right Transformer models on an entire document in which each token is generated only from its prefix context thus involves ``masking out'' part of the attention matrix to prevent each token from attending to its suffix context (essentially, the future). 
Figure \ref{fig:attention} (top) shows the causal attention mechanism used to train a left-to-right language model. 
Figure \ref{fig:attention} describes a simplified Transformer model for both \codegen and our bidirectional language model.
Auto-regressive and left-to-right LMs~\cite{chen2021evaluating, black2021gpt, tunstall2022natural, nijkamp2022codegen} use all previous tokens (i.e., tokens to the left) to predict the probability of the next token (i.e., tokens to the right).
Left-to-right models are useful for program generation tasks, as shown in Figure \ref{code:leftright}.
Specifically, the lower triangular part of the attention matrix remains unmasked (visualized as blue) while attention in the remaining part is masked out (white). This configuration allows each token to attend to itself and all past tokens, but prevents it from attending to future tokens.

Our approach is compatible with any left-to-right language model, but is most effective when the underlying model is large and has been pretrained on a large volume of code data. At present, the \codegen family of models~\cite{nijkamp2022codegen} is most suitable for this role. These are a series of increasingly large left-to-right language models trained for program synthesis in three stages:
\begin{enumerate}
    \item{Each model is first trained on the natural language dataset ThePile, an 825.18 GiB mostly English language text corpus collected by Gao et al.~\cite{gao2020pile} for language modeling. 7.6\% of the dataset is programming language data collected from GitHub repositories with >100 stars.}
    \item{The models are then further trained on data from the Google BigQuery GHArchive dataset, which consists of open-source code across multiple programming languages -- C, C++, Go, Java, JavaScript, and Python.}
    \item{Finally, the models are tuned on the BIGPYTHON dataset, which contains a large amount of Python data.}
\end{enumerate}
Checkpoints after each stage are released for every model size, ranging from 350M to 16B parameters. The 16B model outperforms the original Codex model \cite{chen2021evaluating} on a Python program synthesis task.

While language models are typically used to predict the next token, they can also return the ``hidden'' states from their final Transformer layer. When generating text, these states are converted to a next-token prediction via a simple linear transformation. As such, these states tend to represent the model's knowledge about the evolving context at each point in the file, making them intrinsically useful. As shown in Figure \ref{fig:attention}, we extract the final hidden states for each newline (NL) token in each training sample from CodeGen to produce a condensed sequence representation in which each token represents one line. We pair these with their corresponding location (i.e., line \#5 of a 50 line file) and save these to disk.

To train our model, we load these encoded lines in batches, where we retrieve samples of up to 128 contiguous newline states at a time. We choose this number because the \codegen model can consume a maximum of 2,048 tokens as its input size; inputs with 128 lines almost always fit this token budget. Samples with fewer lines are padded, along with the label vector, to obtain a uniform length. Padding entries are ignored in the loss computation. When files contain more than 128 lines, we sample a series of 128 line windows that cover each faulty line in the file. Specifically, we repeatedly create a sample with up to 128 lines starting from a random offset before the immediate next faulty lines that is not yet covered by a previous segment. We then mark all faulty lines in this segment as covered and repeat until all lines are covered by at least one segment. We choose random starting offsets to ensure that the faulty lines within the split code lines are not consistently at the same indices (e.g., right at the start or in the middle), which would cause our model to memorize certain index locations as faulty lines.  This enables our technique to handle inputs longer than 2048 tokens.

\subsection{Bidirectional Adapter}
\label{sec:adapter}
While left-to-right language models extract rich representations per token, they are ill-suited for fault prediction because the representation of each given token only reflects knowledge of its left-ward context. One solution might predict buggy lines based on the final hidden state, reflecting the model's knowledge after the entire file has been processed, but this creates a bottleneck where that state must store information from each line in the entire file. This bottlenecking phenomenon~\cite{bengio1994learning} is precisely why the NLP field moved away from Recurrent Neural Networks, which represent sequences with a single hidden state, and towards attention-based models, which preserve and use the state of each token~\cite{vaswani2017attention}.

We postulate that we can leverage these rich learned representations at each token by training just a few more Transformer layers that allow the model to exchange information between representations of later and earlier lines, thereby generating a new, bidirectionally aware representation for each line of code. We can do so by removing the causal attention mask that normally prevents the exchange of information with ``future'' tokens in our added layers. In our case, we assume that the entire file has already been written, so this constraint is unnecessary.
This yields a \ul{bidirectional} encoder. As shown in Figure \ref{fig:attention}, the attention masking matrix for the bidirectional model allows all tokens in the sequence to attend to each other (visualized in blue).
We thus train a bidirectional adapter consisting of a configurable number of Transformer layers, following the standard Transformer encoder architecture~\cite{vaswani2017attention}. Concretely, our approach involves five steps, visualized in Figure \ref{fig:attention}:

\textbf{Step 1:} We start with a series of code tokens $C = [c_0, c_1, \ldots, c_N]$. We query a causally pretrained Transformer $\mathcal{T}_{PT}$ to transform these into a representational ``states'' $S \in \mathbb{R}^{N \times D}$, where $D$ represents the pretrained model's dimensionality. This step takes place ``offline'', as we do not tune the pretrained model. 

\textbf{Step 2:} We extract the representations of each newline token to obtain state per line in the original program: $S_{NL} \in \mathbb{R}^{M \times D} = S[c_i=\text{\textbackslash n}]$, where $M$ is the number of original newlines and typically $M \ll N$. We conjecture that these tokens' states reasonably accurately capture the information of their line in the file's prefix context.

\textbf{Step 3:} The dimension of the pretrained model's states, $D$, ranges up to 6,144 for the CodeGen models we built on. We use a significantly smaller dimension $d \ll D$ for our adapter layers, because they are trained on limited data. We first reduce the dimension of $S_{NL}$ to $R_{NL} \in \mathbb{R}^{M \times d} = S_{NL} ~ W_d$ where $W_d \in \mathbb{R}^{D \times d}$ is a learnable weight, equivalent to a fully connected layer. We experiment with dimensions $d \in \{256, 512, 1024\}$

\textbf{Step 4:} We then train an $n$-layer bidirectional Transformer adapter $\mathcal{T}_{A}$ with the same internal dimension $d$. This gives us the final representation of each newline token $A_{NL} \in \mathbb{R}^{M \times d}$, which aims to capture their role in the bidirectional context. We set the number of Transformer layers to $n = 2$.

\textbf{Step 5:} We transform each newline token's representation to a single value ranging from 0 to 1 via a sigmoid-activated dense projection $B = \sigma(R_{NL} ~ W_b)$ where $W_b \in \mathbb{R}^{d \times 1}$. The resulting predictions per newline token can be seen as probability estimates of each line being buggy according to the model. These are compared against the ground-truth labels $T \in \{0, 1\}^M$ using the binary cross-entropy loss $\mathcal{L}_{CE} = T \ln{B} + (1 - T) \ln{(1-B)}$. This loss is backpropagated through all layers up to, but not including, those in the pretrained network to obtain gradients. Given these gradients averaged across a sufficiently large minibatch of samples, the model states are updated to make its predictions more likely to agree with the training labels, using the setup described in Section~\ref{sec:hyperparameters}.

%% file: sections/evaluation.tex
\section{Evaluation}
\label{sec:evaluation}

In this section, we present our approach and results for the following three research questions.

\noindent\textbf{RQ1. How does \tool compare with prior FL techniques?}
We evaluate our technique's performance in comparison with existing FL techniques on the same dataset.

\noindent\textbf{RQ2. How well does \tool's performance generalize to new projects?}
We evaluate \tool's performance on previously-unseen code, to assess its generalizability beyond its training data. 

\noindent\textbf{RQ3. How does each component of \tool impact its performance?}
We conduct an ablation analysis to evaluate the impact of different components on the performance of our model.

\noindent\textbf{RQ4. How generalizable is \tool to other languages and domains?}
We evaluate \tool on different languages and domains.

\subsection{Setup}

\subsubsection{Dataset.}
Our work investigates the effectiveness of LLMs in the setting of fault detection. 
To determine how well our proposed technique can perform on real world faults, we select four datasets with source code and corresponding labeled fault lines. 
\begin{itemize}
\item{\textbf{\defj V1.2.0 }:  A Java benchmark dataset with 395 bugs from 6 Java projects~\cite{just2014defects4j}.} 
We use V1.2.0 for most of our benchmarks instead of the latest version (V2.0.0) to compare on the same dataset as most prior FL techniques.
\item{\textbf{\defj V2.0.0}:  A Java benchmark dataset with additional bugs over \defj V1.2.0~\cite{just2014defects4j}.} To show that our approach can generalize to faults from unseen projects, we further evaluate our tool as trained on \defj V1.2.0 on 226 new bugs from the newest \defj version (from projects totaling 165k more lines of code). We exclude the first 45 bugs in Jsoup and all in Gson/Jacksoncore because of trouble reproducing them (as seen in prior work~\cite{lou2021boosting}).
\item{\textbf{\bpy}: a Python benchmark with 493 bugs from 17 different projects~\cite{widyasari2020bugsinpy}.}
\item{\textbf{\devign:} a C benchmark with 5,260 from two open-source projects~\cite{zhou2019devign}. The original Devign dataset contains 15,512 security vulnerabilities from four different projects~\cite{zhou2019devign}. However, the authors of Devign only released a partial dataset available online.} 

\end{itemize}
\looseness-1
All datasets include fixing commits that correspond to each fault.  We identify faulty statements as those that are changed in the git diff associated with each commit, following prior approaches~\cite{meng2022improving,ray2016naturalness,li2019oopsla}.
We then track line numbers of changed statements as training labels.

\subsubsection{Baselines.}
\tool takes as input source code, and outputs a ranked list of probabilities corresponding to how likely a code line is buggy. To the best of our knowledge, no existing FL approaches take as input only the source code as natural language. However, we compare against existing FL approaches that take as input both source code and test code to observe if an LLM-enabled FL technique can produce comparable results without the dependence on tests or test coverage information.

Our baselines are recent, state-of-the-art statement-level MLFL approaches: DeepFL~\cite{li2019deepfl}, DeepRL4FL~\cite{li2021fault}, and TRANSFER-FL~\cite{meng2022improving}. DeepFL, and DeepRL4FL are MLFL techniques that take the test coverage information as model input. TRANSFER-FL builds on previous test-based MLFL approaches with pretrained information from open-source Java programs. 
We also include Ochiai~\cite{abreu2006evaluation}, the best-performing SBFL approach.  We use the prior techniques' replication packages to compute Top-N scores, including their handling of tied ranks (if any); we follow DeepFL's approach for accounting for tied ranks for Ochiai.  

Our tool produces a fault probability score for each line of a code file (i.e., statement level fault localization). Previous approaches output a ranked list of either suspicious statements or suspicious methods. In particular, DeepFL~\cite{li2019deepfl} is trained at the method level, i.e., predicting which methods are defective. 

To compare, we follow other prior work and use DeepFL’s spectrum and mutation-based features that are applicable to detecting faulty statements.  DeepRL4FL, and TRANSFER-FL perform statement-level fault localization by default, similar to \tool.
Since the repository and processed dataset for DeepRL4FL are not publicly available, we directly cite the experimental results reported in their paper~\cite{li2021fault}. For each of the other compared techniques, we run their tool for a total time of 30 minutes, which is comparable to our tool's training time for 300 epochs.

\subsubsection{Validation. }
For each of our three datasets, we perform a 10-fold cross validation on the entire dataset. Specifically, we shuffle the dataset and train 10 models with 90\% of the training set each, holding out the remaining 10\% for validation, so that each sample in the dataset is held out exactly once.
This is by contrast with some prior evaluations that in their default settings, validate tools within individual projects (using one bug in a given \defj project for validation and training on other bugs in that same project)~\cite{li2019deepfl,lou2021boosting, li2021fault, meng2022improving}.
An effective and robust FL tool using machine learning or language models should be able to predict faulty locations while trained on code from different projects. 

Training FL models on a particular project may produce over-fitting to a particular project and reduces applicability, requiring a relatively rich project and bug history before a technique can be used.  We therefore believe that our 10-fold validation approach is more generalizable for training models on larger code datasets. 
As is done in some prior evaluations~\cite{li2019deepfl,li2021fault}, we also \emph{separately} evaluate the degree to which our model trained on one set of projects generalizes to a set of projects not seen in training (without retraining for those new projects).

We also deploy an early-stopping mechanism for each of our training runs. We checkpoint and record the epoch with the single highest average precision and recall score on the held-out validation dataset after every epoch. Once these scores stop improving for sufficiently many epochs (i.e., around 300 for all our model configurations), we stop training and use the best-performing checkpoint to calculate the Top-N metrics against the ground-truth labels.

\subsubsection{Evaluation Metrics.} We use the following evaluation metrics:

\begin{table}[t]
\caption{\small Hyperparameters used for model training, both for the model trained from scratch and the three models trained on top of the various \codegen models}
\begin{tabular}{l|rrrr}
\toprule
\textbf{Hyperparameter} & \textbf{From Scratch} & \textbf{350M} & \textbf{6B} & \textbf{16B}\\
\midrule
Max learning rate  &  5e-6 & 1e-4 & 7e-6 & 4e-6\\
Min learning rate  &  1e-8 & 1e-7 & 1e-7 & 1e-7\\
Model dimension & 256 & 1024 & 4096 & 6144\\
Layers  &  8 & 2 & 2 & 2\\
Batch size & 64 & 32 & 32 & 32\\
Epochs & 2000 & 300 & 300 & 300\\
\bottomrule
\end{tabular}
\label{table:hyperparams}
\end{table}

\begin{table*}[t]
\caption{\small \tool performance on 395 bugs from \defj V1.2.0, compared to prior techniques (top); on 226 additional bugs from \defj V2.0.0 (middle); and with ablation (bottom, again on defects from \defj V1.2.0)}
\begin{tabular}{l|lrrr}
\toprule
 \textbf{FL type} & \textbf{Technique} & \textbf{Top-1} & \textbf{Top-3} & \textbf{Top-5}   \\
\midrule
\multirow{1}{*}{SBFL}
& Ochiai              & 19 (4.8\%) & 65  (16.5\%) & 99 (25.1\%)  \\
\midrule
\multirow{4}{*}{MLFL}
& DeepFL             & 57 (14.4\%) & 95 (24.1\%) & 135 (34.2\%) \\
& DeepRL4FL          & 71 (18.0\%) & 128 (32.4\%) & 142 (35.9\%) \\
& TRANSFER-FL        & 86 (21.8\%) & 135 (34.2\%) & 160 (40.5\%) \\
\midrule
\multirow{2}{*}{LMFL}
& \tool with \codegen-350M     & 82 (20.8\%) & 106 (26.8\%) & 126 (31.9\%) \\
& \tool with \codegen-6B        & 85 (21.5\%) & 115 (29.1\%) & 160 (40.5\%)\\
& \tool with \codegen-16B       & \textbf{88 (22.3\%)} & \textbf{149 (37.7\%)} & \textbf{183 (46.3\%)} \\
\midrule\midrule
LMFL, new projects & \tool with \codegen-16B & 72 (31.9\%) & 93 (41.2\%) & 123(54.4\%) \\
\midrule\midrule 
\multirow{2}{*}{LMFL Ablation}
& \multicolumn{1}{p{5cm}}{\raggedright $-pretraining$ \\ 
\raggedright (6 layers, trained from scratch)}
& 5 (1.3\%)   & 24 (6.2\%) &  30 (7.6\%)\\
& \multicolumn{1}{p{5cm}}{\raggedright $-bidirectional~adapter$ \\ 
\raggedright (predict directly from \codegen-16B)} & 10 (2.6\%)   & 60 (15.2\%) &  85 (21.5\%)\\
\bottomrule
\end{tabular}
\label{table:topn}
\end{table*}

\emph{Top-N.}
Top-N measures the number of faults with at least one faulty
element located within the first N positions (N=1, 3, 5). Developers only examine a small amount of the most-likely buggy elements within a ranked list~\cite{parnin2011automated}, with particular attention paid to the top-5 elements~\cite{kochhar2016practitioners}. To compare against state-of-the-art techniques, we adopt Top-N following prior work~\cite{li2022fault, li2019deepfl, lou2021boosting}.

\vspace{-0.5em}
\paragraph{AUC of the model's ROC Curve}
Although most developers inspect only top-5 elements in a given list, we also aim to measure how overall prediction compares against the ground truth.
A Receiver Operating Characteristic (ROC) curve shows the performance of one classification model at all thresholds. It can be used to evaluate the overall model strength for making precise and accurate predictions. The area under an ROC curve (AUC) measures the usefulness of a test. AUC is a number between 0 and 1; higher is better.  We measure the AUC at each of our model's top performing points in time, averaging precision and recall. We choose AUC to observe the prediction strength of our models at their peak performance. 

\subsubsection{Ablations}
We conduct an ablation analysis to evaluate the impact of different components on the performance of our model (RQ3). 
We run five variants of our proposed technique for the \defj V1.2.0 dataset.
We first evaluate \tool pretrained on \codegen, and \tool without any pretraining to evaluate the impact of the pretrained large language model's final hidden states.
For the pretrained models, we checkpoint with three different \codegen sizes (i.e., 350 million, 6 billion, and 16 billion parameters) to evaluate the impact of the pretrained model's size on finetuning. 
We also train a version of our model without bidirectional layers, using only the \codegen auto-regressive attention mechanisms for fault localization. We aim to determine if left-to-right LLMs can detect faults directly, without any customization for code understanding.

\subsubsection{Hyperparameters}
\label{sec:hyperparameters}
Table \ref{table:hyperparams} shows the hyperparameters used in training all our models.
We reduced the learning rates until both the training and validation loss converged in a stable manner. Following the established practice in language model training \cite{hoffmann2022training}, we use a learning rate warm-up of 1000 steps and a cosine learning rate decay until a global minimum learning rate of 1e-7 across 20k steps.
Each model is trained on a single GPU. For the \codegen pretrained models, we use a uniform batch size of 32 and perform gradient accumulation to ensure every batch of our data fits on a single GPU.
For a fair comparison of \tool's components (RQ3), we use the same number of  training epochs (300) for all pretrain sizes and projected dimensions. However, the non-pretrained bidirectional model requires a much longer training time (some 2,000 epochs) for the validation loss to converge.

We train all configurations of our model on a uniform dimension of 512, which is projected down from the various \codegen models' hidden state dimensions (i.e., 1024, 4096, and 6144). We use a 8 attention heads for all our models.

\subsubsection{Environment}
All results presented in this section were obtained using an
Intel(R) Xeon(R) 6248R CPU @ 3.00GHz running Debian GNU/Linux 1 and a single Nvidia Quadro RTX 8000 GPU.
Our largest model, \tool with \codegen-16B, takes 20 minutes to train on the \bpy dataset, 30 minutes on the \defj dataset, and 2 hours on the \devign dataset.

\subsection{Results}
\textbf{RQ1: How does \tool compare with prior DL-based FL tools?}
Table \ref{table:topn} (top) details experimental results showing how our tool compares against state-of-the-art FL techniques. The first 4 techniques are from prior approaches; we evaluate our \tool using three \codegen pretrain sizes. The results show the Top-N (N $\in \{1,3,5\}$) score for each technique. 
Table \ref{table:topn} shows that \tool with the largest (16B) pretrained \codegen size outperforms all the compared techniques. Even with smaller pretrain sizes (350M and 6B), \tool performs similarly to the top-performing prior methods.

Per Table \ref{table:topn}, \tool with 16B \codegen pretrain size detects 84 more faults within Top-5 than the top-performing SBFL technique, Ochiai (84.8\% improvement). \tool detects 48 more faults within the Top-5 than the first proposed deep learning based FL technique DeepFL (35.6\% improvement), and 23 more faults within the Top-5 than the latest state-of-the-art test-based MLFL approach TRANSFER-FL (14.4\% improvement). 
For the Top-3 and Top-1 metric, \tool pretrained on the 16B \codegen model can detect 14 more faults (10.4\% improvement) and 2 more faults (2.3\% improvement) than the state-of-the-art tool TRANSFER-FL. We observe that our LMFL technique improves particularly over prior FL techniques when more suspicious lines are inspected (i.e., higher Top-5 scores). 

A Wilcoxon signed-rank test~\cite{woolson2007wilcoxon} indicates that the top-N values the difference between \tool with \codegen-16B and prior techniques in terms of performance at the several top-N values is statistically significant (p-values ranging from 0.01 to 0.03).

When considering Top-1 scores, our approach is only slightly better than TRANSFER-FL, which performs roughly on-par with our \codegen-6B model. 
However, note that prior techniques only achieve comparable results with our tool by requiring readily-available tests and test coverage as input. Writing tests and producing test coverage are time-consuming activities, and tests are not always available or useful when debugging. Furthermore, both DeepFL and TRANSFER-FL techniques include mutation-based fault localization information, which is very time-consuming to collect (i.e., hours of online collection time per bug~\cite{li2019deepfl}).

\begin{tcolorbox}
[colback=white,colframe=black,arc=0pt,boxrule=0.5pt,title=RQ1 Summary,boxsep=2pt,left=1pt,right=1pt,top=1pt,bottom=1pt,fonttitle=\bfseries]
\tool pretrained on the largest \codegen size improves on the state-of-the-art by 14.4\% on Top-5, without relying on test cases, program analysis, or even compilable code.
\end{tcolorbox}

\vspace{0.5em}
\noindent\textbf{RQ2. How well does \tool’s performance generalize to new
projects?}
We additionally evaluate \tool on bugs from the newer \defj V2.0.0, on projects that were not seen in pretraining (an additional 165K lines of code). The ``LMFL, new projects'' row in Table~\ref{table:topn} shows that \tool with 16B \codegen pretrain size detects 72/226 faults in top 1, 93/226 faults in top 3, and 123/226 faults in top 5. 

Although we avoid strong statistical claims in this case study setting, these results are comparable to \tool's performance on projects included in its training data, suggesting that it generalizes well.  Several previously-published techniques are also evaluated for cross-project generalizability, in a variety of experimental settings.  DeepFL and DeepRL4FL repeatedly train a model on N-1 projects and test it on a held-out project; in both cases, performance on the cross-project setting degrades compared to the within-project setting.  GRACE~\cite{lou2021boosting} localizes to the method level (rather than the statement level); its cross-project evaluation also looks at defects from \defj V2.0.0.  GRACE's performance also degrades slightly on  new defects, though less than prior work.  A key advantage of our approach is that \tool generalizes well to unseen projects \emph{without retraining of any kind}.  This argues for our technique's practicality both in terms of performance and time/compute requirements.  

\begin{tcolorbox}
[colback=white,colframe=black,arc=0pt,boxrule=0.5pt,title=RQ2 Summary,boxsep=2pt,left=1pt,right=1pt,top=1pt,bottom=1pt,fonttitle=\bfseries]
\tool pretrained on the largest \codegen size using data from \defj V1.2.0 performs well on bugs in unseen projects (not included in the training data), without additional training costs.
\vspace{0.5em}
\end{tcolorbox}

\begin{table}[t]
\caption{\tool's Top-N Effectiveness on Different Datasets}
\begin{tabular}{l|rrr}
\toprule
\textbf{Metric} &\textbf{\bpy} & \textbf{\defj} & \textbf{\devign}\\
\midrule
\# lines  &  76,672 & 168,960 & 7,180,160\\
Top-1  &  51/493 (10.3\%) & 88/395 (22.3\%) & 1478/5260 (28.1\%)\\
Top-3  &  59/493 (12.0\%) & 149/395 (37.7\%)  & 2050/5260 (39.0\%)\\
Top-5  &  75/493 (15.2\%) & 183/395 (46.3\%)  & 3171/5260 (60.3\%)\\
\bottomrule
\end{tabular}
\label{table:topn-rq4}
\end{table}

\noindent\textbf{RQ3. How does each component of \tool\ impact its performance?} 
The bottom two rows of Table \ref{table:topn} show the impact of pretrained models on \tool's performance. 

\vspace{0.5ex}
\emph{Without Pretraining} We trained our bidirectional language model from scratch, using the same tokenizer as \codegen for tokenizing the inputs. We replace \codegen's token-level representation with a learnable embedding for each token. We then pass these embeddings through 6 bidirectional Transformer layers (a typical minimum for Transformers) and predict the bugginess probability for states corresponding to newline tokens only (other tokens are embedded alongside these but ignored in the final layer). This model, trained on a sample size of 395 (i.e., total number of labeled \defj bugs) can achieve only a Top-1 of 5 (1.3\%), Top-3 of 24 (6.2\%), and Top-5 of 30 (7.6\%). \tool without any pretraining performs significantly worse than \tool based on any size of \codegen. 

\vspace{0.5ex}
\emph{Without the Bidirectional Adapter} We train a single linear projection from \codegen-16B's final hidden states to a bugginess score for each line, thus omitting the bidirectional attention adapter layers. This approach performs better than \tool trained from scratch, with a Top-1 of 10 (2.6\%), Top-3 of 60 (15.2\%), and Top-5 of 85 (21.5\%). This highlights how much program understanding a left-to-right LLM trained on a large corpus of code encodes in its learned representations. Although left-to-right models are not targeted at text-understanding, an LLM that can generate code given a natural language prompt can evidently learn to understand faults to a similar level of top performing SBFL approaches. Given enough fine-tune training on top of the previous task of code generation, CodeGen-16B without any further configuration is able to detect 85 \defj bugs (21.5\%), which is only 14\% worse than the top performing SBFL model Ochiai. However, using \codegen-16B for fault localization directly still performs significantly lower than all \tool models with bidirectional adapter layers.
We perform an additional Wilcoxon signed-rank test~\cite{woolson2007wilcoxon} to observe that the top-N values of \tool with \codegen-16B yields significantly better results than \tool without pretraining and without the bidirectional adapter at $\alpha=0.05$ (p-values of 0.008 and 0.02).

\vspace{0.5ex}
\emph{Underlying LLMs}
Comparing our tool on different pretrained \codegen sizes, we see an improvement in fault detection as the underlying model grows.
\tool pretrained on \codegen-350M improves upon \tool without the bidirectional adapter layers by 72 on Top-1. \tool pretrained on \codegen-6B can detect 3 more faults on Top-1 than \codegen-350M, and \tool pretrained on \codegen-16B can find an additional 3 compared to \codegen-6B.
At higher Top-N targets, the performance improves more steeply with the size of the underlying model. For instance, \tool fine-tuned on \codegen-350M detects 96 more faults than without pretraining, while fine-tuning on top of \codegen-16B uncovers another 153.

\begin{tcolorbox}
[colback=white,colframe=black,arc=0pt,boxrule=0.5pt,title=RQ3 Summary,boxsep=2pt,left=1pt,right=1pt,top=1pt,bottom=1pt,fonttitle=\bfseries]
Although left-to-right language models can directly localize some faults, adding the bidirectional adapter layers is crucial for achieving state-of-the-art fault localization. Furthermore, we show that our tool using the largest pretrained LLM (i.e., \codegen 16B) significantly outperforms all other variations of our model.
\vspace{0.5em}
\end{tcolorbox}

\noindent\textbf{RQ4. How generalizable is \tool to other languages and domains?}
To evaluate our proposed technique on different languages and domains, we run all three pretrain sizes of our tool on the \bpy~\cite{widyasari2020bugsinpy} dataset for localizing Python bugs, and the \devign~\cite{zhou2019devign} dataset for localizing C security vulnerabilities. We believe that measuring our tool on two other languages and one other defect domain can evaluate the effectiveness of modeling code defects as specific behaviors in natural language. 

We observe from Table \ref{table:topn-rq4} that \tool can localize faulty statements with Top-1 of 10.3\% on \bpy, and 28.1\% for \devign. 
We observe that the performance of \tool improves as the size of the training dataset increases. Although \defj has fewer bugs than \bpy, we find that in total, \defj has 53\% more code lines combined from all code files than the \bpy dataset. Since our approach considers source code as natural language, a larger database of code lines gives our models more training data. In particular, our largest dataset \devign with over 7 million lines of code achieves a Top-5 of 60.3\% (i.e., 60.3\% of our model's top-5 suspicious lines have at least one line that is an actual vulnerability). 

\begin{figure}
    \vspace*{-5mm}
\centering
     \begin{subfigure}[t]{0.5\textwidth}
         \centering
         \includegraphics[width=8.25cm]{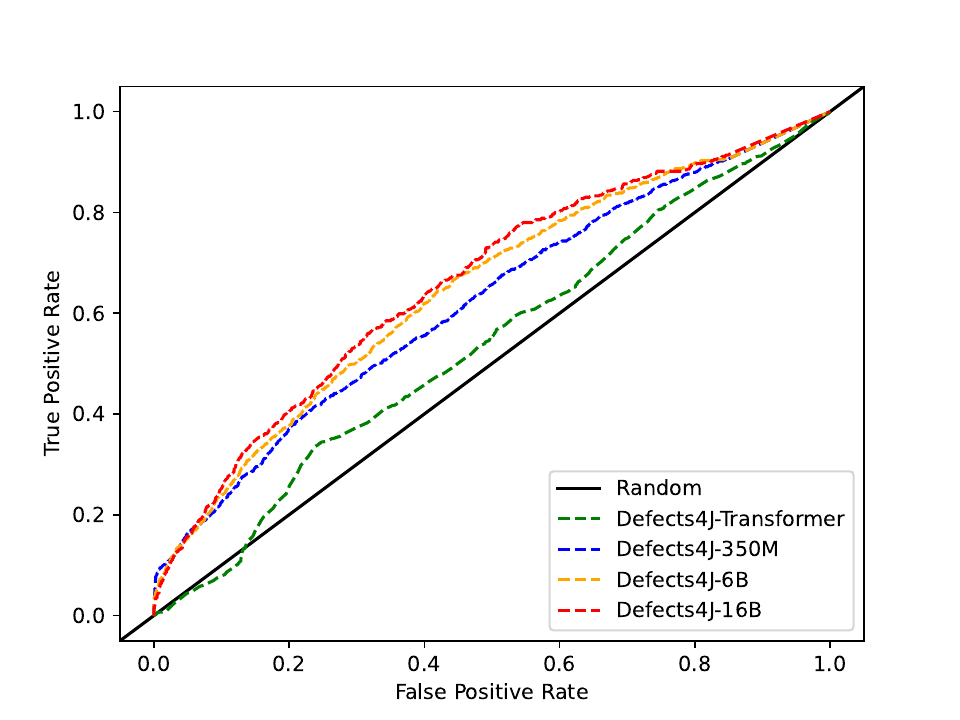}
         \caption{ROC curves on the \defj dataset}
         \label{roc:defects4j}
     \end{subfigure}
     \hfill
     \begin{subfigure}[t]{0.5\textwidth}
         \centering
         \includegraphics[width=8.25cm]{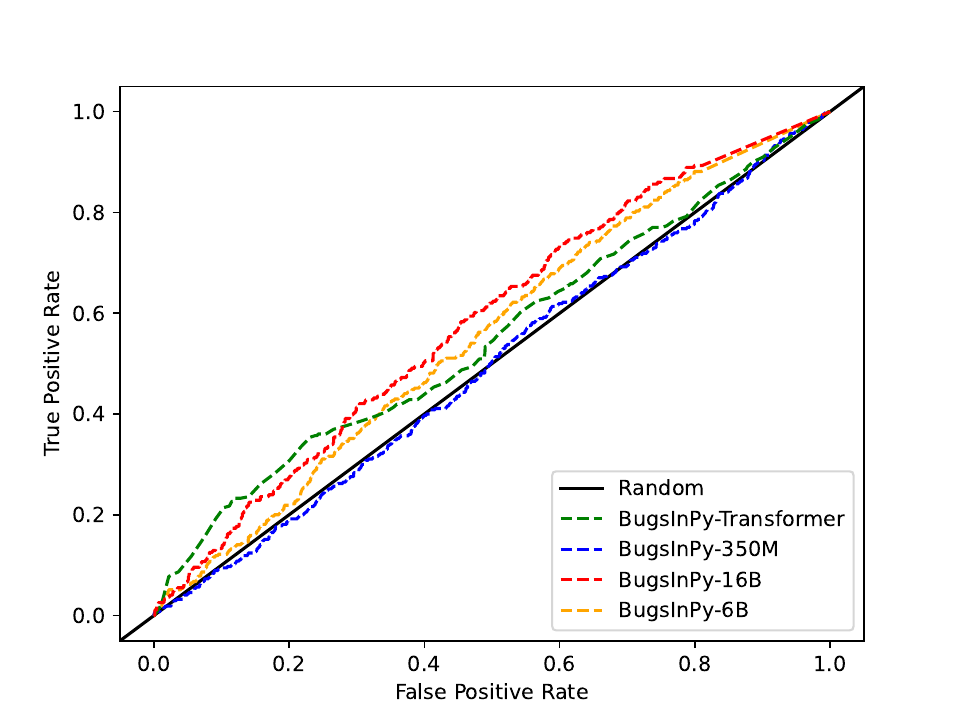}
         \caption{ROC curves on the \bpy dataset}
         \label{roc:bugsinpy}
     \end{subfigure}
     \begin{subfigure}[t]{0.5\textwidth}
         \centering
         \includegraphics[width=8.25cm]{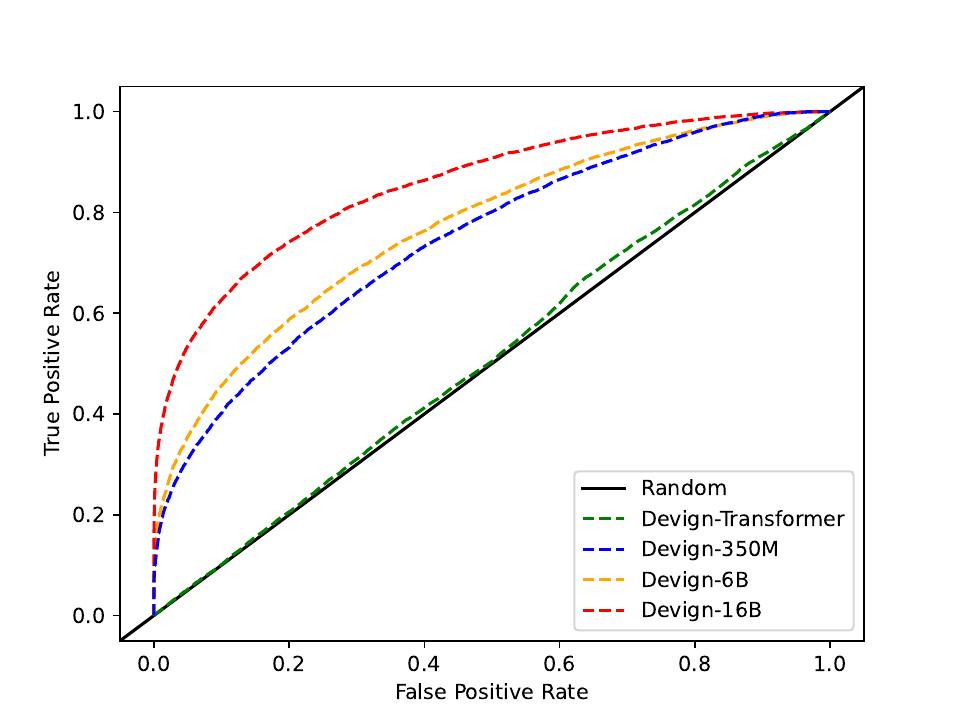}
         \caption{ROC curves on the \devign dataset}
         \label{roc:devign}
     \end{subfigure}
\label{fig:roc}
\vspace{-2mm}
\caption{ROC curves on the completely random prediction, our model without any pretraining (Transformer), and pretrained on \codegen-small (350M), \codegen-medium (6B), and \codegen-large (16B). Higher area under the curve (AUC) represents stronger predictive power.}
\end{figure}

Figures \ref{roc:defects4j}, \ref{roc:bugsinpy} and \ref{roc:devign} show the ROC curve for each of our trained models compared to the completely random curve (i.e., AUC=0.5). 

A ROC curve shows the performance of our model at all classification thresholds. 
The completely random curve has the true positive rate equal to the false positive rate at every classification threshold.
We plot the ROC for our model trained on \defj, \bpy, and \devign after 300 epochs without any pretraining (i.e., the Transformer ROC curve), \codegen-350M pretraining, \codegen-6B pretraining, and finally \codegen-16B pretraining. 

We observe a clear improvement on the AUC as we use the \codegen final hidden states for training, and the AUC continues to improve as we use larger \codegen models. 
In particular, the AUC for Figure \ref{roc:defects4j} yields 0.539 on \defj trained from scratch, 0.573 on \defj trained from \codegen-350M, 0.638 on \defj trained from \codegen-6B, and 0.677 on \defj trained from \codegen-16B.
Figures \ref{roc:bugsinpy} and \ref{roc:devign} show a significant improvement in our model's predictive power as we use a larger dataset of code corpus. \tool with \codegen-16B trained on our smallest dataset \bpy yields an AUC of 0.571, and \tool with \codegen-16B trained on our largest dataset \devign yields an AUC 0.855. We observe that our model's predictive performance on \devign is better than our model's predictive performance on \bpy at all thresholds.

\begin{tcolorbox}
[colback=white,colframe=black,arc=0pt,boxrule=0.5pt,title=RQ4 Summary,boxsep=2pt,left=1pt,right=1pt,top=1pt,bottom=1pt,fonttitle=\bfseries]
Our approach generalizes to other languages and domains, given a large enough labeled. \tool is more confident in its fault detection as the size of both training data and the pretrained model scale up. \tool is also particularly effective for locating security bugs in C where test cases are not available.
\end{tcolorbox}

%% file: sections/related_work.tex
\section{Related Work}
\label{sec:related}
We discuss in the following sections the most recent advances in fault localization and LLM for code.

\subsection{SBFL and MBFL}
Spectrum-based Fault Localization (SBFL)~\cite{abreu2006evaluation, abreu2007accuracy, wong2007effective, zhang2011localizing} and Mutation-based Fault Localization (MBFL)~\cite{moon2014ask, musco2017large, papadakis2015metallaxis, zhang2010test}  have been extensively studied for fault localization. 

SBFL calculates the suspiciousness score of each code line to represent the probability of the line being faulty. SBFL measures the number of failed and passed tests that cover each code line to generate the suspiciousness score. To generate suspiciousness scores, SBFL uses a ranking formulae based on the test coverage information of each code line.
Although SBFL is widely accepted due to its simplicity and efficiency, SBFL takes coverage as the only input information, and more specifically only retrieves the number of tests from test coverage information. Test coverage alone cannot always encapsulate the faulty behaviors from code lines.

MBFL techniques~\cite{moon2014ask, musco2017large, papadakis2015metallaxis, zhang2010test} mitigate the limitations of SBFL by applying mutation testing to generate mutants for the original program under a test suite. The original program is mutated with one syntactic change that corresponds to a predefined rule. The mutation rules are called mutation operators, (e.g., change if $(a$==$b)$ into if $(a$!=$b)$). MBFL techniques use mutants to check the impacts of code elements on the test outcomes for fault localization.
MBFL techniques consider the impact information whereas SBFL does not. However, MBFL techniques would fail in cases where an element does not have a possible mutant for impact simulation. 

Although leveraging test coverage and program mutation is relatively simpler than billion parameter deep learning models, SBFL cannot accurately rank the statements with the same spectrum scores, and MBFL fails in situations where a mutant cannot be instantiated. Furthermore, both SBFL and MBFL depend on the extent the test suite can soundly and completely cover the source code. Our work does not depend on test coverage, but instead uses the naturalness of code~\cite{ray2016naturalness}. 

\vspace{0.5em}
\subsection{MLFL}
Machine learning fault localization techniques have been proposed using insights from program analysis on code behavior. Prior MLFL techniques train on data such as test coverage~\cite{briand2007using, zhang2017deep}, co-changing method declaration and corresponding statement-level calls~\cite{li2022fault}, or the program's code structure, such as the abstract syntax tree (AST)~\cite{li2021fault}.  
Recent deep learning techniques, such as GRACE~\cite{lou2021boosting} and FixLocator ~\cite{li2022fault}, encode the AST and test coverage as graph representations and learn to rank faulty methods with graph neural networks (GNN)~\cite{scarselli2008graph}. GNNs can directly analyze graph structured information with all topological dependencies reserved, so as to not lose information through data preprocessing. DeepRL4FL ~\cite{li2021fault} use Convolution
Neural Network (CNN) \cite{krizhevsky2017imagenet} applied on code coverage (CC) matrix.
TRANSFER-FL~\cite{meng2022improving} leverages the deep semantic features and transferred knowledge from open-source data to improve fault localization. DeepFL~\cite{li2019deepfl} and TRANSFER-FL~\cite{meng2022improving} combine semantic-based, spectrum-based, and mutation-based features and use a multi-layer perceptron (MLP) model for fault localization.

In contrast, \tool does not require test code, an AST parser, and includes a text tokenizer and embedding layer by construction. 
Prior MLFL models base their architecture on recurrent or convolutional neural networks. \tool leverages attention mechanisms~\cite{vaswani2017attention} and build bidirectional adapter layers on pretrained left-to-right LLMs directly on source code.

\subsection{LMs for Code}
Language models (LMs) are widely applied to natural language~\cite{bengio2000neural}. Recent advances in language modeling using code as training data have shown that LMs can perform code completion~\cite{desai2016program} and generate code based on natural language~\cite{raychev2014code} with impressive results. Large Language Models (LLMs) have drastically raised performance on these tasks \cite{chen2021evaluating}.
However, studies have shown that LLM code generation techniques, such as Codex~\cite{chen2021evaluating} or GPT-Neo~\cite{black2021gpt}, can be prompted to generate buggy programs, including ones with security vulnerabilities~\cite{pearce2021empirical}. Furthermore, many existing code based LLMs are not publicly available for customization for specific tasks~\cite{xu2022systematic}. Our work shows how to build a bidirectional language model fine-tuned on a left-to-right model, and trained specifically for the task of fault localization.

%% file: sections/threats.tex
\section{Discussion and Threats}
\emph{Why does it work?}
Recent LLMs train on such a large corpus of data that they can generate functionally correct code bodies from simple natural language documentation \cite{chen2021evaluating}.

Models such as Codex \cite{chen2021evaluating} and InCoder \cite{fried2022incoder} can perform the opposite direction as well: generate natural language docstring from code snippets alone. These abilities suggest that LLMs extract a significant amount of semantic knowledge from the code they process, even as their objective is just to predict each next token.

We believe that this ability to reason about code semantics translates naturally to reasoning about defects or vulnerabilities. While the model may have been mostly trained on correct code, it likely notes surprising, bug-related patterns much as traditional language models do~\cite{ray2016naturalness}, incorporating this information into token representation, since knowledge of a potential mistake is important for next-token prediction as well.
Our first attempt to directly train a fault localizer on top of an LLM specialized in program synthesis supported this notion by yielding surprisingly strong results (Top-5 score of 85), but still fell short of common baselines.
Our second key observation is that the LLM's knowledge at any given token is also incomplete, lacking awareness of suffix. Many bugs, including the one in Figure~\ref{code:lang47-code}, cannot be reliably determined until close to the end of the program, so the model is incentivized to store information about potentially important missing knowledge (in the  example: whether \texttt{getNullText()} can return \texttt{null}) in the representation of earlier tokens (those on line 5) for later tokens to consider.
Our adapter layers were subsequently introduced to exploit the observation that the representations of later tokens might contain valuable information for determining the bugginess of earlier ones. The resulting model outperformed all baselines and showed strong signs of improving with the scale of the underlying LLM, strongly supporting this approach.

\vspace{-0.5em}
\paragraph{Threats to validity}
One threat to \textit{internal validity} lies in using \texttt{diff} descriptions of bug-fixing commits to identify faulty statements.  Different annotators can disagree about the true cause of a defect~\cite{dbgbench}, and parsing commits provides a noisy proxy for truly faulty lines associated with a bug.  We mitigate this threat first 
by relying on well-established previously-published datasets with historical bug fixes.  
These datasets are manually curated to confirm that the commits in question do fix a  given bug, and \defj's bug-fixing commits are further pruned to include only bug-relevant changes (reducing the influence of unrelated ``tangled'' changes such as refactorings). 
We note that \defj, on which the bulk of our experiments are performed, is a very common dataset in prior fault localization work, supporting comparison and consistency.  The use of the commit to indicate ground-truth faulty statements or methods has similar precedent in the literature (e.g., but not limited to, refs~\cite{meng2022improving,ray2016naturalness,li2019oopsla}).
At worst, a developer fix provides a conservative approximation of code defectiveness.  We further mitigate the risks of mistakes in our implementation by releasing our scripts, code, and data as part of a replication package for this work available at \url{https://github.com/squaresLab/LLMAO}.
Threats to \textit{external validity} lie in whether results on our benchmarks will generalize to real-world contexts. To reduce this threat, we evaluate on the widely-used \defj-V1.2.0 \cite{just2014defects4j} with hundreds of real-world bugs. Although it is a widely-used benchmark, recent techniques may be overfitting to it~\cite{durieux2019empirical}; we therefore additionally train our tool on two other benchmarks, \bpy and \devign. 

As \tool is trained on top of \codegen, which takes as training data roughly 65 GiB of code from GitHub repositories up to 2021, we can not fully mitigate the bias that our datasets could be included in its training data. However, \codegen was trained on a very large volume of code, meaning that the model retains relatively little memory of our partition of that training data. The training data of \codegen also does not include the manually annotated labels for bugs, but rather the repositories of the faulty code directly, which would not bias our fault localization models.

Threats to \textit{construct validity} lie in measurements used. We use multiple metrics widely used to evaluate both fault localization and ML models. We also perform our experiments under 10-fold cross validation and across three datasets to strengthen generalizability. 

%% file: sections/conclusion.tex
\section{Conclusions}
\label{sec:conc}
In this paper, we propose \tool, an LLM-based approach for localizing program defects, which include general logic defects as well as security vulnerabilities. We perform an empirical study on 395 real bugs from \defj, 493 bugs from \bpy, and 5,260 security vulnerabilities from \devign. Our results show that \tool can outperform existing state-of-the-art deep learning based fault localization techniques without the use of insights from extensive program analysis, or any test cases. 
In particular, \tool can localize  48/395 more faults within the Top-5 than the first proposed deep learning based fault localizer, DeepFL, which is guided by SBFL and MBFL artifacts that require extensive manual labor to attain. \tool can localize 23/155 more bugs within Top-5 than TRANSFER-FL, which is the latest state-of-the-art deep learning based fault localizer.
The comparison of AUC on different versions of our model shows that training on top of larger LLMs improves performance significantly and that our approach based on bidirectional adapter layers is essential for achieving state-of-the-art localization scores. The experimental results show that pretraining on the largest \codegen model (e.g., 16 billion parameters) achieves the highest AUC on all our studied datasets.
To the best of our knowledge, \tool is the first DL based tool to localize security vulnerabilities on a line level without requiring test cases or even compilable code.